\documentclass{article}

    \PassOptionsToPackage{numbers, compress}{natbib}

\usepackage[final, nonatbib]{neurips_2019}
\usepackage[utf8]{inputenc} 
\usepackage[T1]{fontenc}    
\usepackage{hyperref}       
\usepackage{url}            
\usepackage{booktabs}       
\usepackage{amsfonts}       
\usepackage{nicefrac}       
\usepackage{microtype}      
\usepackage{graphicx}
\usepackage{caption}
\usepackage{subcaption}
\usepackage{authblk}
\usepackage{lipsum}

\title{Validation of a deep learning mammography model in a population with low screening rates}

\author[1]{Kevin Wu}
\author[1]{Eric Wu}
\author[2]{Yaping Wu}
\author[2]{Hongna Tan}
\author[1]{Greg Sorensen}
\author[2]{Meiyun Wang}
\author[1]{Bill Lotter}
\affil[1]{DeepHealth, Inc.}
\affil[2]{Henan Provincial People's Hospital}

\begin{document}

\maketitle
\begin{abstract}
    
A key promise of AI applications in healthcare is in increasing access to quality medical care in under-served populations and emerging markets. However, deep learning models are often only trained on data from advantaged populations that have the infrastructure and resources required for large-scale data collection. 
In this paper, we aim to empirically investigate the potential impact of such biases on breast cancer detection in mammograms.
We specifically explore how a deep learning algorithm trained on screening mammograms from the US and UK generalizes to mammograms collected at a hospital in China, where screening is not widely implemented.
For the evaluation, we use a top-scoring model developed for the Digital Mammography DREAM Challenge.
Despite the change in institution and population composition, we find that the model generalizes well, exhibiting similar performance to that achieved in the DREAM Challenge, even when controlling for tumor size. We also illustrate a simple but effective method for filtering predictions based on model variance, which can be particularly useful for deployment in new settings. While there are many components in developing a clinically effective system, these results represent a promising step towards increasing access to life-saving screening mammography in populations where screening rates are currently low.
\end{abstract}

\section*{Introduction}

In the United States, screening mammography has been attributed as a major factor in averting hundreds of thousands of breast cancer deaths~\cite{hendrick2019breast}. The prevalence of screening mammography, however, is often dramatically lower in many emerging markets, including China~\cite{fan2014breast, song2015breast}.  Inadequate funding and a shortage of qualified readers are contributing factors to China’s low screening rates~\cite{fan2014breast}. 
AI could increase access to screening via its scalability and potential to provide low cost and accurate initial interpretation for many mammograms.
For instance, a triage-style clinical implementation could significantly reduce overall workload for clinicians, while maintaining high diagnostic accuracy~\cite{yala2019deep}. In this scenario, the majority of women with mammograms confidently interpreted as normal by the AI software could be recommended for subsequent routine mammography the following year, where the remainder of women would be referred for further clinician interpretation and work-up, thus lightening the clinical load.

There has indeed been much recent work illustrating the promise of AI in mammogram interpretation~\cite{carneiro2015unregistered, lotter2017multi, ribli2018detecting, morrell2018large, kooi2017large, wu2019deep}. However, it is often unclear how well these methods generalize, as they are usually evaluated on data drawn from the same distribution as the training set, i.e. similar populations and/or hospitals~\cite{kim2019design}. With goals of deploying such solutions in different markets, their performance on the intended population must be verified, especially as it is often infeasible to collect enough training data from these populations. 
Generalization is certainly not guaranteed given the high sensitivity of deep learning models to input statistics~\cite{zech2018variable, pan2019generalizable}. For mammography in particular, there are known (and possibly unknown) biological differences in mammograms between US and Chinese populations, such as a greater proportion of dense breasts in the Chinese population~\cite{bae2016breast}, which can make mammogram interpretation more difficult. Here, we tested how well a top-performing model from Digital Mammography DREAM Challenge~\cite{dream}, a large-scale data science competition that concluded in early 2018, generalizes to a collection of mammograms acquired at a Chinese hospital. 
Additionally, we explore the use of an empirical measure of prediction uncertainty, quantified as the variance over models and data augmentations in an ensemble, as a prediction deferment strategy, which could be useful in deployment in novel settings.

\section*{Methods}
\subsubsection*{Model Architecture and Training Procedure}
\begin{figure}
    \centering
    \includegraphics[scale=0.3]{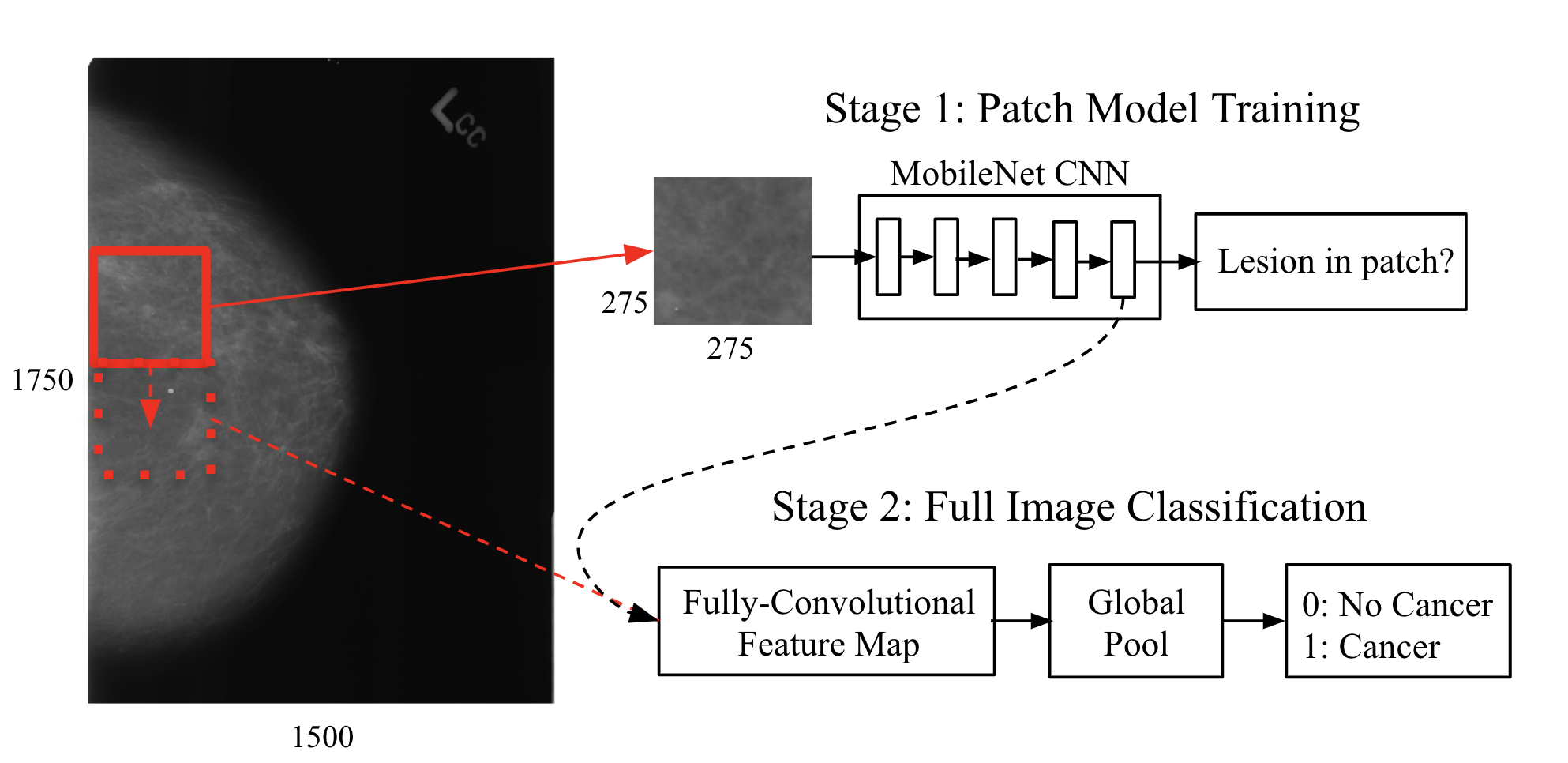}
    \caption{Illustration of the two-stage approach used to develop our top-scoring Dream Challenge model, which was trained on US and UK data, and evaluated on data from a Chinese hospital in this work.}
    \label{fig:model}
\end{figure}

The tested algorithm was designed for the Digital Mammography DREAM Challenge~\cite{dream}, a unique competition where neither the training data nor the testing data (all from the United States) were available for download. Instead, competitors made submissions using Docker containers, which would have access to the data when run on the DREAM servers. Another challenge was that only breast-level labels were available for training; specifically there were only binary cancer/no-cancer labels for each patient and laterality (left/right), with no tumor localization information. Given the “needle-in-a-haystack” nature of detecting cancer in mammograms, training standard image classification approaches using only image-level labels quickly leads to overfitting. To more effectively train on the DREAM data, we developed a two-stage training scheme, which first consisted of training convolutional neural networks (CNNs) on cropped mammogram patches using publicly-available, strongly-labeled data; and then using these CNNs to initialize a fully-convolutional network that outputs a classification given a full-scale image~\cite{lotter2017multi} (Figure ~\ref{fig:model}). By first training to predict the presence of a cancerous lesion in a 256x256px mammogram patch, the CNN is better initialized for end-to-end training on full-scale images, for which we resize to have a height of 1750px. We used the DDSM~\cite{heath2000digital} and Optimam~\cite{optimam} datasets for patch pre-training. 
DDSM consists of 2,602 scanned film mammography studies from the US (35\% cancers, 33\% benigns, and 31\% normals), while the version of Optimam used here contains 13,973 digital mammography studies from the UK (26\% cancers, 2.5\% benigns, and 72\% normals).

For the backbone of our network, we used MobileNet~\cite{howard2017mobilenets}, as its memory efficient structure enabled full-image training on the GPUs available in the DREAM Challenge. Converting the patch model to the full-image model consisted of using a global average pooling layer on top of the final convolutional feature map layer, followed by a single fully-connected layer. The overall training procedure consisted of 1) patch-level training on DDSM \& Optimam, 2) image-level training on DDSM \& Optimam, and 3) image-level training on the DREAM dataset. Due to the class imbalance between cancer/no-cancer images, we sampled equally from each class during all three training stages. 
The final model consisted of an ensemble of three models trained in a similar fashion. Each model averages scores from vertically mirrored orientations of the same image, and an image-level score is computed as a weighted average between the three models. A study-level score is computed by averaging image-level scores across views (cranio-caudal (CC) and mediolateral oblique (MLO)), before taking the max score over lateralities (left and right). This submission achieved an area under the receiver operating characteristic curve (AUROC) of 0.90, which was the highest submission score over all phases of the Challenge.

\subsubsection*{Evaluation on Data from a Chinese Hospital}
To test the generalization of the DREAM model, an evaluation dataset consisting of 2533 cases (533 pathologically-proven cancers, 1000 pathologically-proven benigns, and 1000 normals) was gathered from an urban hospital in China. Each case consisted of the four standard mammographic views (CC and MLO of both the left and right breast). The data was retrospectively collected from a contiguous period of time from 2012-2017. Given low screening rates, the data came from diagnostic exams, i.e. exams where the patient presented with symptoms, so the distribution of tumor sizes from the cancer cases contained more large tumors (64\% larger than 2cm) than would be expected in a typical United States screening population~\cite{tumorsize}. Thus, we report results on the original data distribution, as well as a tumor-size normalized performance, using bootstrap re-sampling simulations to approximately match the tumor size distribution observed in US screening data~\cite{nmd}. All results were calculated by running the containerized DREAM submission locally at the Chinese hospital with no transfer of imaging data. 
As a measure of model uncertainty for a given image~\cite{ayhan2018test}, we quantify the variance in prediction scores across the three models and two vertical orientations (6 total scores). An uncertainty score per breast is then calculated as the average in variances across views for that breast.

\section*{Results}
\begin{figure}
    \centering
    \includegraphics[scale=0.5]{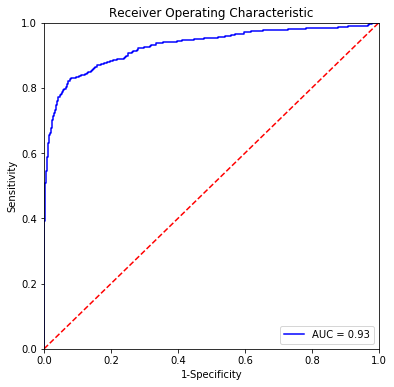}
    \caption{Receiver operating characteristic (ROC) curve of the DREAM model evaluated on the tested Chinese dataset containing 2533 total cases.} 
    \label{fig:roc_curve}
\end{figure}

On the original Chinese dataset, the DREAM model achieved $0.93 \pm 0.01$ AUROC for breast-level predictions, as illustrated in Figure~\ref{fig:roc_curve}.
In terms of a simulated triage scenario, these results would translate to interpreting 60\% of normal mammograms as such, while operating at 95\% sensitivity. 
When controlling for tumor size using 1000 bootstrap simulations to match US screening statistics, the model achieved $0.90 \pm 0.03$ AUROC. 
This level of performance is thus consistent with the $0.90$ AUROC results from the DREAM Challenge.
Estimating breast density using an open source tool~\cite{oncoserve}, we observe an AUROC of $0.914$ on dense breasts and $0.946$ non-dense breasts, respectively.
The difference of 0.03 AUROC is also consistent with results from the DREAM Challenge.
Finally, as an additional performance measure, the model achieves an AUROC of $0.94$ when only considering normal and malignant cases (i.e., excluding the benigns).

\begin{figure}[h]
    \centering
    \includegraphics[width=0.495\textwidth]{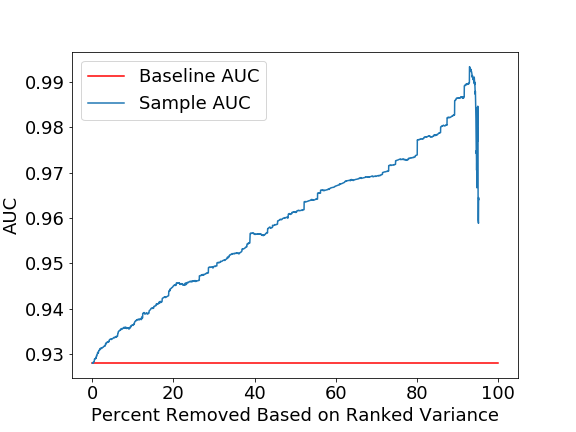}
    \includegraphics[width=0.495\textwidth]{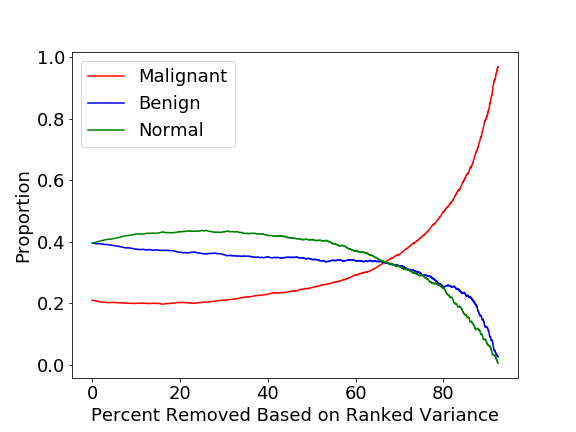}
    \caption{(Left) AUROC by percentage of samples filtered based on the uncertainty criteria. We compute and rank order the uncertainty criteria for each sample in the evaluation set. The x-axis is the percentage of data removed, starting with highest variance samples, while the y-axis reports the AUROC achieved on that subset of data. (Right) Relative proportion of malignant, benign, and normal cases in the data as the percent removed based on ranked variance increases. The plot infers that benign lesions generally show highest model variance, followed by normal and malignant lesions.} 
    \label{fig:filtering}
\end{figure}

Additionally, we find that the variance between prediction scores in the model ensemble (3 models across 2 image orientations) provides a practical empirical method of filtering model predictions.
The cases that exhibit high prediction variance seem to also be the cases that the algorithm tends to misclassify, as excluding cases based on this metric tends to increase AUROC (Fig.~\ref{fig:filtering}). 
In analyzing the proportion of each type of case (normal, benign, malignant) that is filtered as the percentage of filtered cases increases, we observe that there is a slight trend towards initially excluding benign cases (Fig.~\ref{fig:filtering}), though the relative proportion of each type remains relatively consistent up to excluding \textasciitilde40\% of cases.

\section*{Discussion}
While the development of better performing models will always be a goal in AI research, validating that existing algorithms can perform equally well in new clinical settings is crucial to ensure that AI can effectively serve broad populations. In this study, we demonstrated that a model that was trained on data from the US and UK and achieved state-of-the-art performance on the Digital Mammography DREAM Challenge generalizes well to data collected at an urban Chinese hospital. 
While the testing data used in this study largely came from diagnostic exams, as screening rates are relatively low in China, the performance was comparable to the United States results even when sampling to approximately match tumor size statistics found in the US.
We also note that, while the data was diagnostic, each case consisted of the four standard mammographic (screening) views.
The multi-faceted data used to train our model, consisting of film and digital mammograms and three datasets in total, could be a factor in the demonstrated generalizability. It could also be the case that the variation in mammographic image statistics within a demographic population tends to be larger than the variability between demographics.
Since the evaluation performed here is without additional training, it is also reasonable to expect performance improvements in a setting where new data is available for fine-tuning the model.
In addition to demonstrating generalizability, we propose a method for the deferring of model predictions via an empirical uncertainty measure formulated as the variance of scores between models in an ensemble. 
This can be especially useful in novel settings, where cases exhibiting a level of uncertainty above a threshold can be reserved solely for physician interpretation. Overall, while further work is needed, these results are a promising initial step towards the deployment of a mammography AI system to a population where screening mammography is currently limited.

\small

\bibliographystyle{unsrt}
\bibliography{references.bib}

\end{document}